\documentclass{article} 
\usepackage{arxiv_fmt,times}
\usepackage{hyperref}
\usepackage{url}
\usepackage[square,numbers]{natbib}
\usepackage{ulem}

\usepackage{amsmath}
\usepackage{amssymb}
\usepackage{algorithmic}
\usepackage{graphicx}

\def\E{\mathbb{E}}
\def\weight{\bar\omega}
\def\Wbar{\overline{W}}

\providecommand{\U}[1]{\protect\rule{.1in}{.1in}}

\usepackage{listings}
\usepackage{balance}

\lstnewenvironment{code}[2]{\lstset{caption=#1,label=#2}}{}

\title{Asynchronous Anytime Sequential Monte Carlo}

\author{
Brooks Paige \hspace{.5cm} Frank Wood \\
Department of Engineering Science \\
University of Oxford \\
Oxford, UK \\
\texttt{\{brooks,fwood\}@robots.ox.ac.uk} \\
\And
Arnaud Doucet \hspace{.5cm}   Yee Whye Teh\\
Department of Statistics \\
University of Oxford \\
Oxford, UK \\
\texttt{\{doucet,y.w.teh\}@stats.ox.ac.uk} 
}

\nipsfinalcopy 

\begin{document}

\maketitle

\begin{abstract}

We introduce a new sequential Monte Carlo algorithm we call the \textit{particle cascade}.
The particle cascade is an asynchronous, anytime alternative to traditional particle filtering algorithms.
It uses no barrier synchronizations which leads to improved particle throughput and memory efficiency.
It is an anytime algorithm in the sense that it can be run forever to emit an unbounded number of particles while keeping within a fixed memory budget.  We prove that the particle cascade is an unbiased marginal likelihood estimator which means that it can be straightforwardly plugged into existing pseudomarginal methods.
\end{abstract}

\section{Introduction}
\label{sec:introduction}

Particle filter based inference techniques require blocking barrier synchronization at resampling steps which limits parallel throughput and is costly in terms of memory.  
We introduce a new asynchronous particle filter algorithm that has statistical efficiency competitive with standard resampling algorithms, and has sufficiently higher particle throughput such that it is, on balance, more efficient per unit time.
The approach uses locally-computed decision rules for each particle that do not require block synchronization of all particles,
instead only requiring sharing summary statistics with particles that follow.  
In our algorithm each resampling point acts as a queue rather than a barrier: each particle chooses the number of its own offspring using by comparing
its own weight to the weights of particles which previously reached the queue, updates its own weight, then proceeds without waiting.


An anytime algorithm is an algorithm which can be run continuously, generating progressively better solutions when afforded additional computation time.
Traditional particle filter (PF) algorithms are not anytime in nature; all particles need to be propagated in lock-step to completion in order to compute expectations.   
Once a particle set runs to termination, inference cannot straightforwardly be continued by simply doing more computation.  
The naive strategy of running sequential Monte Carlo (SMC) again and merging the resulting sets of particles is suboptimal due to bias (see \cite{whiteley2013} for explanation).  
More complex methods (i.e. particle Metropolis Hastings and iterated conditional sequential Monte Carlo (iCSMC) \cite{andrieu2010particle}) for correctly merging particle sets produced by additional SMC runs are closer to anytime in nature but suffer from burstiness as big sets of particles are computed then emitted at once and, fundamentally, the inner-SMC loop of such algorithms  still suffers the kind of excessive synchronization performance penalty that the particle cascade directly avoids.
Our asynchronous SMC algorithm, the \textit{particle cascade}, is anytime in nature. 
The particle cascade can be run indefinitely, without resorting to merging of particle sets,
and with only a fixed (tunable) memory requirement.

%

\subsection{Related work}
\label{sec:relatedwork}

Our algorithm shares a superficial similarity to Bernoulli branching numbers \cite{crisan1999filtering} and other search and exploration methods which have been used for particle filtering, where each particle samples some number of children to propagate to the next observation.
Like the particle cascade (and in contrast to most traditional resampling algorithms), the total number of particles which exist at each generation is allowed to gradually increase and decrease.
However, computing branching correction numbers is also generally a synchronous operation, requiring all particle weights to be known at each observation in order to choose an appropriate number of offspring; this also precludes use of an anytime algorithm.

Parallelizing the resampling step of sequential Monte Carlo methods has drawn increasing recent interest as the effort progresses to scale up algorithms to take advantage of high-performance computing systems and GPUs.
A recent approach to removing the global collective resampling operation, quite different from the particle cascade method introduced here, can be found at \cite{murray2014parallel}.

Another recent method for running arbitrarily many particles within a fixed memory budget introduced in \cite{jun14icml} focuses on keeping track of random seeds used to generate proposals, allowing particular particles to be deterministically ``replayed''; 
a first pass through all particles computes the normalizing constant of the particle weights, and a second pass re-executes those which are chosen to continue to the next generation.
However, the algorithm as presented there still relies on a synchronous resampling step, and lacks the anytime property of our approach.

\section{Background}
\label{sec:background}

We begin by briefly reviewing particle filtering as generally formulated on state-space models.
Suppose we have a non-Markovian dynamical system with latent random variables $X_0,\ldots,X_N$ and observed random variables $Y_0,\ldots,Y_N$
described by the joint density
\begin{align}
p(X_n|X_{0:n-1},Y_{0:n-1}) &= f(X_n|X_{0:n-1}) \nonumber \\
p(Y_n|X_{0:n},Y_{0:n-1}) &= g(Y_n|X_{0:n}),
\label{eq:joint-density}
\end{align}
where $X_0$ is drawn from some initial distribution $\mu(\cdot)$, 
and $f$ and $g$ are conditional densities.

Given observed values $Y_{0:N}=y_{0:N}$, we approximate the posterior distribution $p(X_{0:n}|y_{0:n})$
with a weighted set of $K$ particles, with each particle $k$ denoted $x^k_{0:n}$ for $k = 1,\dots,K$. 
%
Particles are propagated forward from proposal densities $q(x_n|x_{0:n-1})$ and re-weighted at each observation $n=1,\ldots,N$:
\begin{align}
x^k_n|x^k_{n-1} &\sim q(\cdot|x^k_{n-1}) \\
w^k_n &= \frac{f(x^k_n|x^k_{0:n-1})g(y_n|x^k_{0:n})}{q(x^k_n|x^k_{0:n-1})} \\
W^k_n &= W^k_{n-1} w^k_n
\end{align}
where $w^k_n$ is the weight associated with observation $y_n$ and $W^k_n$ is the weight of particle $k$ after observation $n$.
We assume that exact evaluation of $p(x_{0:N}|y_{0:N})$ is intractable, requiring only that the conditional terms likelihoods $g(y_n|x^k_{0:n})$ can be evaluated pointwise.
In many complex dynamical systems, or in black-box simulation models, evaluation of $f(x^k_n|x^k_{0:n-1})$ may be prohibitively costly or even impossible.
As long as we are capable of simulating from the system we can set our proposal distribution $q(\cdot) \equiv f(\cdot)$, in which case
the particle weights are simply $w^k_n = g(y_n|x^k_{0:n})$, eliminating the need to compute the conditional densities $f(\cdot)$ directly.

By normalizing the weights $W^k_n$, defining 
\begin{align}
\weight^k_n = \frac{W^k_n}{\sum_{j=1}^K W^j_n},
\end{align}
we can approximate the posterior distribution $p(X_{0:N}|y_{0:N})$ with a weighted set of $K$ particles
\begin{align}
p(X_{0:N}|y_{0:N}) &\approx  \sum_{k=1}^K \weight^k_N \delta_{x^k_{0:N}}(X_{0:N}).
\end{align}
In the very simple sequential importance sampling setup described here, the marginal likelihood can be estimated by 
\begin{align}
\hat p(y_{0:n}) = \frac{1}{K} \sum_{k=1}^K W^k_n.
\label{eq:ml0}
\end{align}


\subsection{Resampling, degeneracy, synchronization}

The algorithm described above suffers from a degeneracy problem wherein the normalized weights $\weight^1_n, \dots, \weight^K_n$ become mostly very close to zero for even moderately large $n$.
Traditionally this is combated by introducing a resampling step: as we progress from $n$ to $n+1$, particles with high weights are duplicated and particles with low weights are discarded.
Many difference schemes for resampling particles exist; see \cite{douc05comparisonof} for an overview, with discussion and theoretical results for several common approaches.
One can think of a resampling scheme as a method for drawing the number of offspring particles $M^k_{n+1}$ that each particle $k$ will produce after stage $n$.
After resampling, all outgoing particles from $n$ to $n+1$ receive a new outgoing weight $V^k_{n+1}$, and we have 
\begin{align}W^k_{n+1} = V^k_{n+1}w^k_{n+1}.
\end{align}
In most traditional resampling schemes the outgoing weights of all $K$ particles are deterministically set to be equal, i.e.~with $V^k_{n+1} = 1/K$;
a valid resampling scheme then must satisfy the unbiasedness condition
\begin{align}
\E[M^k_{n+1}] = K\weight^k_n.
\end{align}
Introducing this resampling step prevents all the probability mass in our approximation to the posterior from accumulating on a single particle.
In this version of the algorithm, where a resampling step is added at every $n$, the marginal likelihood can be estimated by 
\begin{align}
\hat p(y_n|y_{0:n-1}) = \frac{1}{K} \sum_{k=1}^K w^k_n;
\label{eq:ml1}
\end{align}
it is well-known that the estimate of the marginal likelihood is unbiased \cite{delmoral04feynman}.

\subsection{Limitations}

Our goal is to scale up to very large numbers of particles, using a parallel computing architecture where each particle is simulated as a separate process or thread.
In order to resample at each $n$ we must compute the normalized weights $\weight^k_n$, requiring us to wait until all individual particles have both finished forward simulation and computed their individual weight $W^k_n$ before any can proceed.
While the forward simulation itself is trivially parallelizable, the weight normalization and resampling step is a synchronous, collective operation.
In practice this can lead to significant underuse of computing resources in a multiprocessor environment,
hindering our ability to scale up to large numbers of particles.

Memory limitations on finite computing hardware also limit the number of simultaneous particles $K$ we are capable of running in practice.
All $K$ particles must move through the system together, and all must exist simultaneously; if the total memory requirements of $K$ particles is greater than the available system RAM, then a substantial overhead will be incurred from regularly swapping memory contents to disk.

\section{The Particle Cascade}
\label{sec:methods}

The particle cascade algorithm we introduce addresses both these limitations: it does not require synchronization, and keeps only a bounded number of particles alive in the system at any given time.
Instead of resampling, we will consider particle branching, where each particle can result in 0 or more offspring. 
These branching events happen asynchronously and mutually exclusively, i.e.~they are processed one at a time.

\subsection{Local branching decisions}

At each stage $n$ of the particle filter, particles process observation $y_n$. 
Without loss of generality, we can define an ordering on the particles $1,2,\ldots$ in the order they arrive at $y_n$. 
This order need not be independent of the state of the particles $x^k_{0:n}$.

We keep track of the running average weight $\Wbar^k_n$ of the first $k$ particles to arrive at observation $y_n$ in an online manner:
\begin{align}
\Wbar^k_n &= W^{k}_n &&\text{for $k=1$,}\\
\Wbar^k_n &= \frac{k-1}{k} \Wbar^{k-1}_n + \frac{1}{k} W^{k}_n 
&&\text{for $k=2,3,\ldots.$}
\end{align}
The number of children of particle $k$ should depend on the weight $W^k_n$ of particle $k$ relative to those of other particles.  
Particles with higher relative weight are more likely to be located in a high posterior probability part of the space, and should be allowed to spawn more child particles.

In the online asynchronous particle system as described here, we do not have access to the weights of future particles when processing $k$.  
Instead we will compare $W^k_n$ to the current average weight $\Wbar^k_n$ among particles processed thus far. 
Specifically, the number of children, which we denote by $M^k_{n+1}$, will depend on the ratio
\begin{align}
R^k_n = \frac{W^k_n}{\Wbar^k_n}.
\end{align}
Each child of particle $k$ will be assigned a weight $V^k_{n+1}$ such that the total weight of all children $M^{k}_{n+1}V^{k}_{n+1}$ has expectation $W^k_n$.

There is a great deal of flexibility available in designing a scheme for choosing the number of child particles; we need only be careful to set $V^k_{n+1}$ appropriately.
Informally, we would like $M^k_{n+1}$ to be large when $R^k_n$ is large.
If $M^k_{n+1}$ is sampled in such a way that $\E[M^k_{n+1}] = R^k_n$, then we set the outgoing weight $V^k_{n+1}=\Wbar^k_n$.  
Alternatively, if we are using a scheme which deterministically guarantees $M^k_{n+1}>0$, then we set $V^k_{n+1}={W}^k_n/M^k_{n+1}$. 

A simple approach would be to sample $M^k_{n+1}$ independently conditioned on the weights.
In such schemes we could draw each $M^k_{n+1}$ from some simple distribution, e.g.~a Poisson distribution with mean $R^k_n$, or a discrete distribution over the integers $\{ \lfloor R^k_n\rfloor, \lceil R^k_n \rceil \}$.
However, one issue that arises in such approaches
where the number of children for each particle is conditionally independent, is that the variance of the total number of particles at each generation can grow without bound. 
Suppose we start the system with $K_0$ particles.  The number of particles at subsequent stages $n$ is given recursively as
$K_n = \sum_{k=1}^{K_{n-1}} M^k_n$. 
We would like to avoid situations in which the number of particles becomes too large, or collapses to 1.



Instead, we will allow $M^k_n$ to depend on the number of children of previous particles at $n$, in such a way that we can stabilize the total number of particles in each generation.  
Suppose that we wish for the number of particles to be stabilized around $K_0$.  After $k-1$ particles have been processed, we expect the total number of children produced at that point to be approximately $k-1$, so that if the number is less than $k-1$ we should allow particle $k$ to produce more children, and vice versa.  Similarly, if we already currently have more than $K$ children, we should allow particle $k$ to produce less children.  
We use a simple scheme which satisfies these criteria, where the number of particles is chosen at random when $R^k_n < 1$, and set deterministically when $R^k_n \geq 1$:
\begin{align}
(M^k_{n+1},V^k_{n+1}) = \begin{cases}
(0,0) \text{ w.p. $1-R^k_n$, }
	&\text{if $R^k_n<1$;} \\
(1,\Wbar^k_n) \text{ w.p. $R^k_n$, } 
	&\text{if $R^k_n<1$;} \\
(\lfloor R^k_n\rfloor,\frac{W^k_n}{\lfloor R^k_n\rfloor})
	&\text{if $R^k_n\ge 1$ and $\sum_{j=1}^{k-1} M^j_{n+1}> \min(K_0,k-1)$;} \\
(\lceil R^k_n\rceil,\frac{W^k_n}{\lceil R^k_n\rceil})
	&\text{if $R^k_n\ge 1$ and $\sum_{j=1}^{k-1} M^j_{n+1}\le \min(K_0,k-1)$.}
\end{cases}
\label{eq:resampling_rule}
\end{align}

We pause here to take note of the anytime nature of this algorithm --- any given particle passing through the system needs only the previous weights $\Wbar^k_n$ in order to make its local branching decisions, not the previous particles themselves.
Thus it is possible to run this algorithm for some fixed number of initial particles $K_0$, inspect the output of the $K_N$ completed particles which have left the system,
and then decide whether to continue inference by initializing additional particles.

%

\subsection{Computing expectations and marginal likelihoods}

Samples drawn from the particle cascade can be used to compute expectations in the same manner as samples from a standard particle filter; 
that is, given some function $\varphi(\cdot)$, we normalize weights $\weight^k_n = \frac{W^k_n}{\sum_{j=1}^{K_n} W^j_n}$ analogously to before and approximate the posterior expectation by
\begin{align}
\E_{X_{0:N}|Y_{0:N}}[\varphi(X_{0:N})]
&\approx
 \sum_{k=1}^{K_N} \weight^k_N \varphi(x^k_{0:N}).
\end{align}

We can also use the particle cascade to define an estimator of the marginal likelihood $p(y_{0:n})$,
\begin{align}
\hat p(y_{0:n}) &= \frac{1}{K_0} \sum_{k=1}^{K_n} W^k_n.
\label{eq:ml-estimator}
\end{align}
The form of this estimate is fairly distinct from the standard SMC estimators 
in Section~\ref{sec:background}.
In terms of predictive densities,
one can think of $\hat{p}\left(y_{0:n}\right)  $ as
\begin{align}
\hat{p}\left(  y_{0:n}\right)  =\hat{p}\left(  y_{0}\right)
{\prod\limits_{i=1}^{n}}
\hat{p}\left(  y_{i}| y_{0:i-1}\right)
\end{align}
where
\begin{align}
\hat{p}\left(  y_{0}\right)   &= \frac{1}{K_{0}}\sum_{k=1}^{K_{0}}%
W_{0}^{k}, &
\hat{p}\left(  \left.  y_{n}\right\vert y_{0:n-1}\right)   &=\frac
{\sum_{k=1}^{K_{n}}W_{n}^{k}}{\sum_{k=1}^{K_{n-1}}W_{n-1}^{k}}\text{ for
}n\geq1.
\end{align}
It is interesting to note that the incrementally updated $\Wbar^k_n$ statistics in the denominator of $R^k_n$ are very directly tied to the marginal likelihood estimate; that is,
$\hat p(y_{0:n}) = \frac{K_n}{K_0} \Wbar^k_n$.


\subsection{Theoretical properties, unbiasedness}
\label{sec:theoretical}

Here we show that the marginal likelihood estimator $\hat p(y_{0:n})$ defined in Eq.~\ref{eq:ml-estimator} is unbiased; i.e.~$\E[\hat p(y_{0:n})] = p(y_{0:n})$.
In a very general form, we can summarize the particle cascade algorithm as
\begin{itemize}
\item Initialization at $n=0$: for $k=1,...,K_{0}$ sample $X_{0}^{k,0}\sim
\mu(  \cdot)  $ and compute $W_{0}^{k}=g( y_{0}| X_{0}^{k,0})  .$
\item At each time $n\geq0$, perform
\begin{enumerate}
\item Resampling step: resample $\left\{  W_{n}^{k},X_{0:n}^{k,n}\right\}
_{k=1}^{K_{n}}$ to obtain $\left\{  \widetilde{W}_{n}^{k},X_{0:n}%
^{k,n+1}\right\}  _{k=1}^{K_{n+1}}$
\item Forward simulation step: for $k=1,...,K_{n+1}$ sample $X_{n+1}^{k,n+1}\sim
f\left(  \left.  \cdot\right\vert X_{0:n}^{k,n+1}\right)  $, and set
$W_{n+1}^{k}=\widetilde{W}_{n}^{k}g\left(  \left.  y_{n+1}\right\vert
X_{0:n+1}^{k,n+1},y_{0:n}\right)  $ and $n\leftarrow n+1.$
\end{enumerate}
\end{itemize}
We denote by $B(E)$ the space of bounded real-valued functions on a space $E$. 
We make
the following assumption on the resampling step.

\textbf{Assumption R}. For any $n\geq0,$ we have $p(  K_{n}>0)
=1$ and for any $\varphi\in B(  \mathcal{X}^{n})  $
\begin{align}
\mathbb{E}\left[  \left.  \sum_{k=1}^{K_{n+1}}\widetilde{W}_{n}^{k}\text{
}\varphi\left(  X_{0:n}^{k,n+1}\right)  \right\vert \mathcal{F}_{n}\right]
=\sum_{k=1}^{K_{n}}W_{n}^{k}\text{ }\varphi\left(  X_{0:n}^{k,n}\right)
\end{align}
where $\mathcal{F}_{n}$ denotes the natural filtration associated to all the
random variables generated by the particle algorithm before resampling at time $n$.
We also denote by $\widetilde{\mathcal{F}}_{n}$ denotes the natural filtration
associated to all the random variables generated by the particle algorithm just after the
resampling step at time $n$.

The resampling step of the particle cascade corresponds to
\begin{align}
\mathbb{E}\left[  \left.  \sum_{k=1}^{K_{n+1}}\widetilde{W}_{n}^{k}\text{
}\varphi\left(  X_{0:n}^{k,n+1}\right)  \right\vert \mathcal{F}_{n}\right]
&=\mathbb{E}\left[  \left.  \sum_{k=1}^{K_{n}}M_{n+1}^{k}V_{n+1}^{k}\text{
}\varphi\left(  X_{0:n}^{k,n}\right)  \right\vert \mathcal{F}_{n}\right]
=\sum_{k=1}^{K_{n}}W_{n}^{k}\text{ }\varphi\left(  X_{0:n}^{k,n}\right)
,\label{eq:unbiased}%
\end{align}
i.e. each particle $X_{0:n}^{k,n}$ has $M_{n+1}^{k}$ 
 offspring of
associated weight $V_{n+1}^{k}$ so that $K_{n+1}=\sum_{k=1}^{K_{n}}M_{n+1}%
^{k}.$


\textbf{Proposition 1}. Assume Assumption R holds and $g(
y_{n}| \cdot,y_{0:n-1})  :\mathcal{X}^{n-1}\rightarrow
\mathbb{R}$ is in $B(  \mathcal{X}^{n-1})  $ for any $n=0,...,N$. Then we
have
\begin{align}
\mathbb{E}[  \hat{p}(  y_{0:n}) ]  =p(
y_{0:n})  .
\end{align}

\textbf{Proof of Proposition 1}. The proof follows from a backward induction.
 We have%
 \begin{align*}
 \mathbb{E}\left[  \hat{p}\left(  y_{0:n}\right)  \right]    &
 =\mathbb{E}\left[  \mathbb{E}\left[  \left.  \frac{1}{K_{0}}\sum_{k=1}^{K_{n}%
 }W_{n}^{k}\right\vert \widetilde{\mathcal{F}}_{n-1}\right]  \right]  \\
 & =\mathbb{E}\left[  \frac{1}{K_{0}}\sum_{k=1}^{K_{n}}\widetilde{W}_{n-1}%
 ^{k}\underset{p\left(  \left.  y_{n}\right\vert X_{0:n-1}^{k,n},y_{0:n-1}%
 \right)  }{\underbrace{\int f\left(  \left.  x_{n}\right\vert
 X_{0:n-1}^{k,n}\right)  g\left(  \left.  y_{n}\right\vert X_{0:n-1}%
 ^{k,n},x_{n},y_{0:n-1}\right)  dx_{n}}}\right]  \\
 & =\mathbb{E}\left[  \mathbb{E}\left[  \left.  \frac{1}{K_{0}}\sum
 _{k=1}^{K_{n}}\widetilde{W}_{n-1}^{k}p\left(  \left.  y_{n}\right\vert
 X_{0:n-1}^{k,n},y_{0:n-1}\right)  \right\vert \mathcal{F}_{n-1}\right]
 \right]  \\
 & =\mathbb{E}\left[  \mathbb{E}\left[  \left.  \frac{1}{K_{0}}\sum
 _{k=1}^{K_{n-1}}W_{n-1}^{k}p\left(  \left.  y_{n}\right\vert X_{0:n-1}%
 ^{k,n-1},y_{0:n-1}\right)  \right\vert \widetilde{\mathcal{F}}_{n-2}\right]
 \right]  \\
 & =\mathbb{E}\left[  \mathbb{E}\left[  \left.  \frac{1}{K_{0}}\sum
 _{k=1}^{K_{n-1}}\widetilde{W}_{n-2}^{k}p\left(  \left.  y_{n-1:n}\right\vert
 X_{0:n-2}^{k,n-1},y_{0:n-2}\right)  \right\vert \mathcal{F}_{n-2}\right]
 \right]  \\
 & =\mathbb{E}\left[  \mathbb{E}\left[  \left.  \frac{1}{K_{0}}\sum
 _{k=1}^{K_{n-2}}W_{n-2}^{k}p\left(  \left.  y_{n-1:n}\right\vert
 X_{0:n-2}^{k,n-2},y_{0:n-2}\right)  \right\vert \mathcal{F}_{n-2}\right]
 \right]  \\
 & =\mathbb{E}\left[  \frac{1}{K_{0}}\sum_{k=1}^{K_{0}}W_{0}^{k}p\left(
 \left.  y_{1:n}\right\vert X_{0}^{k,0},y_{0}\right)  \right]  \\
 & =p(  y_{0:n})  .
 \end{align*}

\section{Active bounding of memory usage}
\label{sec:implementation}

In an idealized computational environment, with infinite available memory, our implementation of the particle cascade could begin by launching (a very large number) $K_0$
particles simultaneously which then gradually propagate forward through the system.
In practice, only some finite number of particles, probably much smaller than $K_0$, can be simultaneously simulated efficiently.
Furthermore, the initial particles are not truly launched all at once, but rather in a sequence, introducing a dependency in the order in which particles arrive at each observation $n$.

\begin{figure}[tb]
\centering
\includegraphics[width=0.49\textwidth]{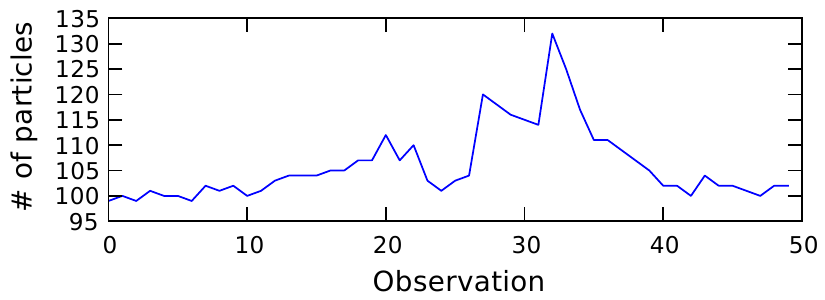}
\includegraphics[width=0.49\textwidth]{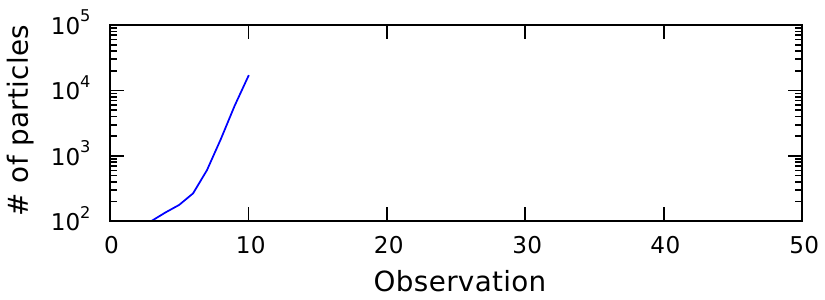}
\caption{
Number of particles $K_n$ at each of $n = 0, \dots, 49$ for a one-dimensional linear Gaussian model, initialized with 100 particles.
(left) When the order of the particles arriving at each $n$ is subject to a random permutation, then the number of particles is reasonably stable, staying at or near 100.
(right) When the order of the particles arriving at each $n$ is completely deterministic, then the total number of particles quickly explodes, in this case exceeding 15000 by $n=11$.
}
\label{fig:particle_instability}
\end{figure}

While the resampling scheme in Eq.~\ref{eq:resampling_rule} is designed to stabilize the number of particles over time, we can still see an explosion in the number of particles $K_n$.
The degree to which the particle count becomes unstable depends on the extent to which the ordering of the particles is permuted as we progress to each $n$.
In Fig.~\ref{fig:particle_instability} we compare a best-case situation where the ordering of particles at $n$ is completely independent of the ordering of particles at $n+1$,
to a worst-case situation where the ordering of particles is completely preserved from $n$ to $n+1$.
In practice, a na\"ive implementation of the incremental resampling scheme will have a very strong dependence in ordering across $n$ --- a particle which is one of the first to reach stage $n$ is quite likely one of the first to reach stage $n+1$ as well.

Our implementation of the particle cascade addresses these issues by explicitly injecting randomness into the execution order of particles, and by imposing a machine-dependent hard cap on the number of simultaneous extant processes.
This permits us to run our particle filter system indefinitely, for arbitrarily large initial particle counts $K_0$, while consuming only a fixed computational budget.

Each particle in our implementation runs as an independent operating system process.
In order to efficiently run a large number of particles, we impose a hard limit
limit $\rho$ on the total number of particles which can simultaneously exist in the particle system;
most of these will generally be sleeping processes.
The ideal choice for this number will vary based on hardware capabilities, but in general should be made as large as possible.

Scheduling across particles is managed via a global \textit{first-in random-out} process queue of length $\rho$; this can equivalently be conceptualized as a random-weight priority queue.
Each particle corresponds to a single live process, augmented by a single additional control process which is responsible only for spawning additional initial particles (i.e.~incrementing the initial particle count $K_0$).
When any particle $k$ arrives at any likelihood evaluation $n$, it computes its target number of child particles $M_{n+1}^k$ and outgoing particle weight $V_{n+1}^k$.
If $M_{n+1}^k = 0$ it immediately terminates; otherwise it enters the queue.
Once this particle either enters the queue or terminates, some other process continues execution --- this process is chosen uniformly at random, and as such may be a sleeping particle at any stage $n < N$, or it may instead be the control process which then launches a brand new particle.
At any given time, there are some number of particles $K_\rho < \rho$ currently in the queue, and so the probability of resuming any particular individual particle, or of launching a new particle, is $\frac{1}{K_\rho + 1}$.
If the particle released from the queue has exactly one child to spawn, it advances to the next observation and repeats the resampling process. 
If, however, a particle has more than one child particle to spawn, rather than launching all child particles at once it launches a single particle to simulate forward, decrements the total number of particles left to launch by one, and itself re-enters the queue.

In the event that the process count is fully saturated (i.e.~the process queue is full), then we forcibly prevent particles from duplicating themselves and creating new children.
If we release a particle from the queue which seeks to launch $m > 1$ additional particles when the queue is full, we instead collapse all the remaining particles into a single particle; this single particle represents a ``virtual'' set of particles, but does not actually create a new process and requires no additional CPU or memory resources.
We keep track of a particle count multiplier $C^k_n$ that we propagate forward along with the particle.
All particles are initialized with $C^k_0 = 1$, and then when a particle collapse takes place, update their multiplier at $n+1$ to $m C^k_n$.

This affects the way in which running weight averages are computed; suppose a new particle $k$ arrives with multiplier $C^k_n$ and weight $W^k_n$.
We incorporate all these values into the average weight immediately, and update $\overline{W}^{k}_n$ taking into account the multiplicity, with
\begin{align}
\overline{W}^{k}_n &= \frac{k-1}{k + C^k_n - 1} \overline{W}^{k-1}_n + \frac{C^k_n}{k + C^k_n - 1} W^{k}_n
&&\text{for $k=2,3,\ldots$.}
\end{align}
This does not affect the computation of the ratio $R^k_n$.
We preserve the particle multiplier, until we reach the final $n = N$; then, after all forward simulation is complete, we re-incorporate the particle multiplicity when reporting the final particle weight $W^k_N = C^k_N V^k_N w^k_N$.
The system is initialized by seeding the system with a number of initial particles $\rho_0 < \rho$ at $n=0$,
creating $\rho_0$ active initial processes.

The ideal choice for the process count constraint $\rho$ may vary across operating systems and hardware configurations; 
online optimization of this parameter remains an avenue for future work. 



\section{Experiments}
\label{sec:experiments}


\begin{figure}[t]
\begin{center}
\includegraphics[width=.49\textwidth]{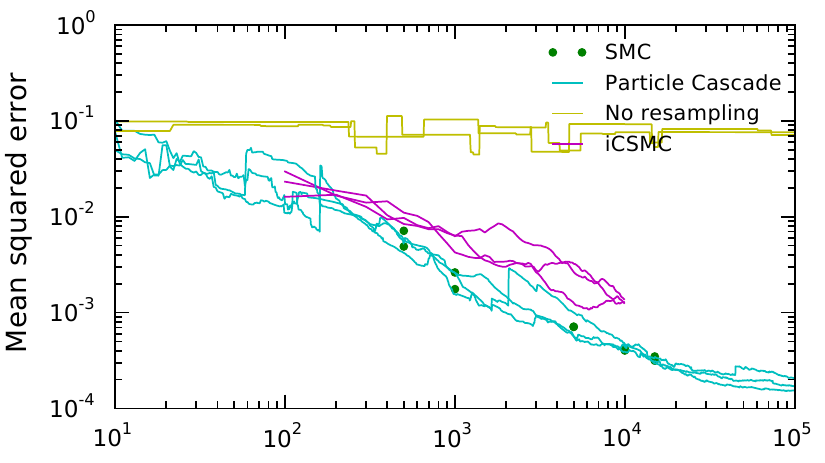}
\includegraphics[width=.49\textwidth]{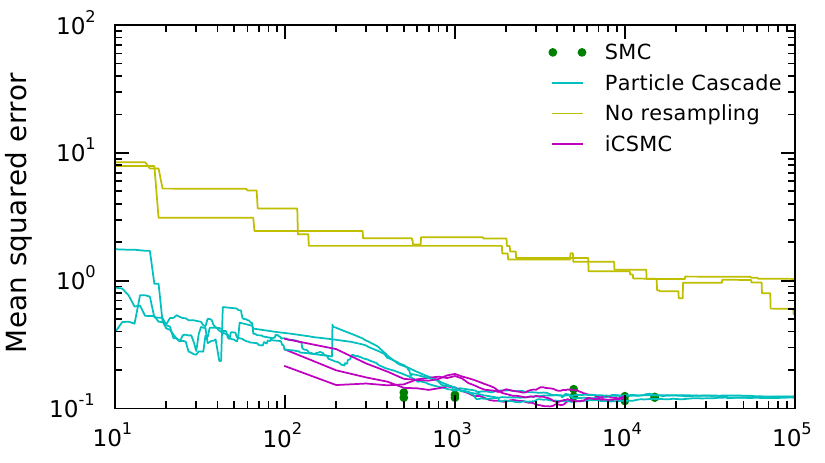}
\includegraphics[width=.49\textwidth]{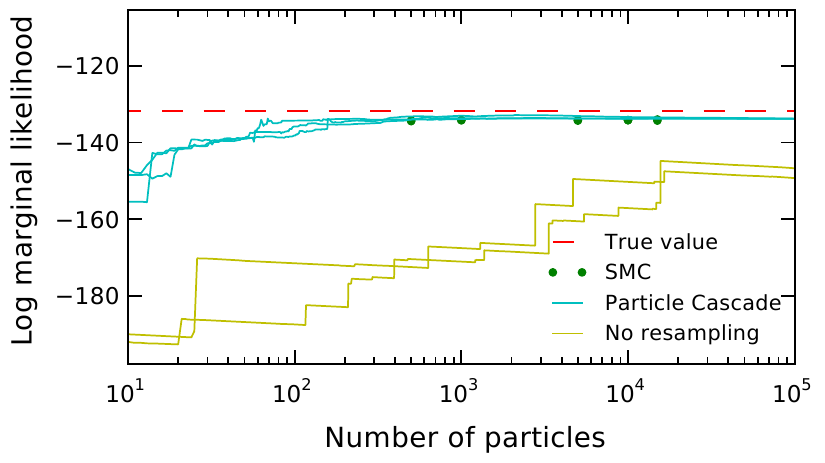}
\includegraphics[width=.49\textwidth]{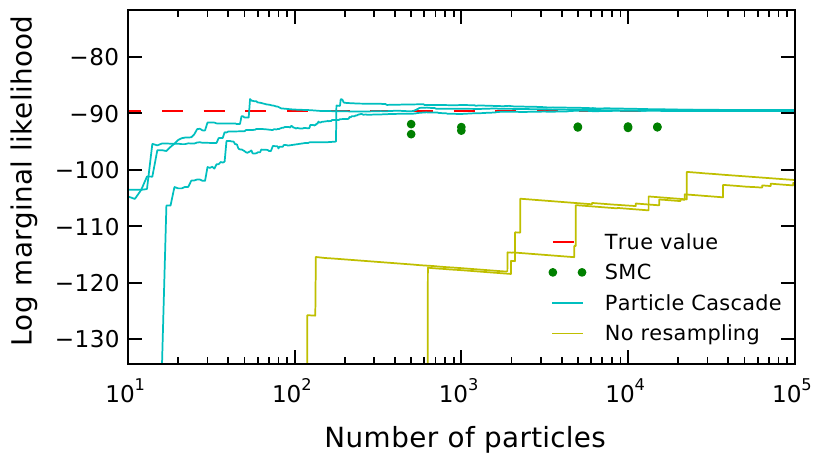}
\caption{All results are reported over multiple independent replications, shown here as independent lines.
(top) Convergence of estimates to ground truth vs.~number of particles, shown as
(left) MSE of marginal probabilities of being in each state for every observation $n$ in the HMM, and
(right) MSE of the latent expected position in the linear Gaussian state space model.
(bottom) Convergence of marginal likelihood estimates to the ground truth value (marked by a red dashed line), for (left) the HMM, and (right) the linear Gaussian model.
}
\label{fig:converge-to-ground-truth}
\end{center}
\vspace{-.5em}
\end{figure}


We run a preliminary set of experiments on two simple state space models, each with $N=50$ observations, with the goal of demonstrating the overall validity and utility of the particle cascade algorithm.
Results are presented here on two simple models.
The first is a hidden Markov model (HMM) with 10 latent discrete states, each with an associated Gaussian emission distribution;
the second is a one-dimensional linear Gaussian model.
In both models we can use an exact algorithm to compute posterior marginals at each $n$ and compute the marginal likelihood $Z = p(y_{1:N})$.

These experiments are not designed to stress-test the particle cascade; rather, they are designed to show that performance of the particle branching scheme closely approximates that of the fully synchronous particle filter, even in a small-data small-complexity regime where we expect particle filter performance to be very good.
In addition to comparing to a particle filter which resamples synchronously, we also compare to a worst-case particle filter in which we never resample, instead propagating particles forward deterministically with a single child particle at every $n$.
While the statistical (per-sample) efficiency of this approach is quite poor, it is fully parallelizable with no blocking operations in the algorithm at all, and thus provides a ceiling estimate of the raw sampling speed attainable in our overall implementation.

We also benchmark against what we believed to be the most practically competitive similar approach, iterated conditional SMC \cite{andrieu2010particle}.
Iterated conditional SMC corresponds to the particle Gibbs algorithm in the case where parameter values are known; by using a particle filter sweep as a step within a larger MCMC algorithm, iCSMC provides a statistically valid approach to sampling from a posterior distribution by repeatedly running sequential Monte Carlo sweeps each with a fixed number of particles.
One downside to iCSMC is that it does not provide an estimate of the marginal likelihood.

On both these models we see the statistical efficiency of the particle cascade is approximately in line with the true particle filter, slightly outperforming the iCSMC algorithm and significantly outperforming the fully parallelized non-resampling approach.
This suggests that the approximations made by computing weights at each $n$ based on only the previously observed particles, and the total particle count limit imposed by $\rho$, do not have an adverse effect on overall performance.
In Fig.~\ref{fig:converge-to-ground-truth} we plot convergence per particle to the true posterior distribution, as well as convergence in our estimate of the normalizing constant.


\subsection{Performance and scalability}

\begin{figure}[t]
\begin{center}
\includegraphics[width=.49\textwidth]{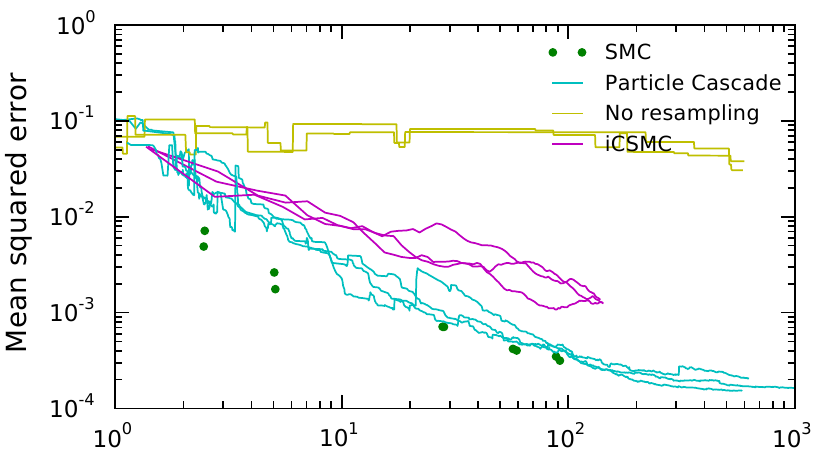}
\includegraphics[width=.49\textwidth]{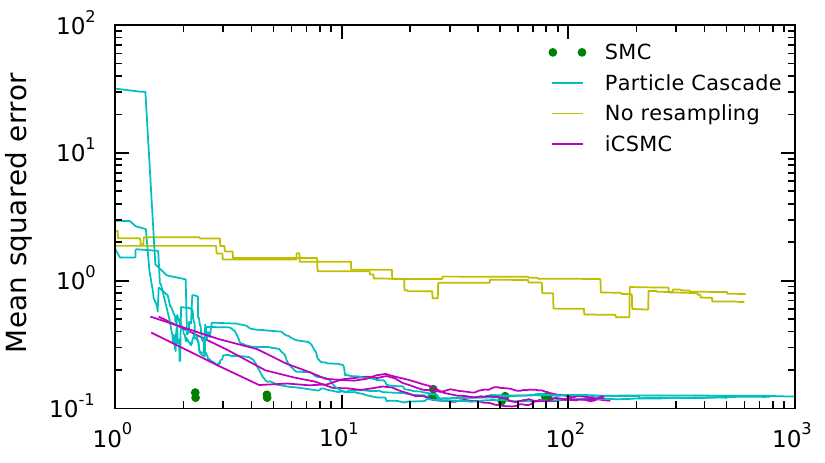}
\includegraphics[width=.49\textwidth]{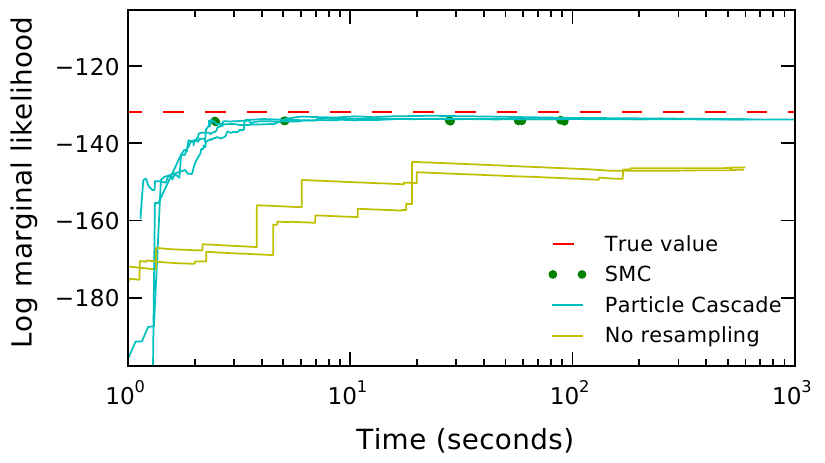}
\includegraphics[width=.49\textwidth]{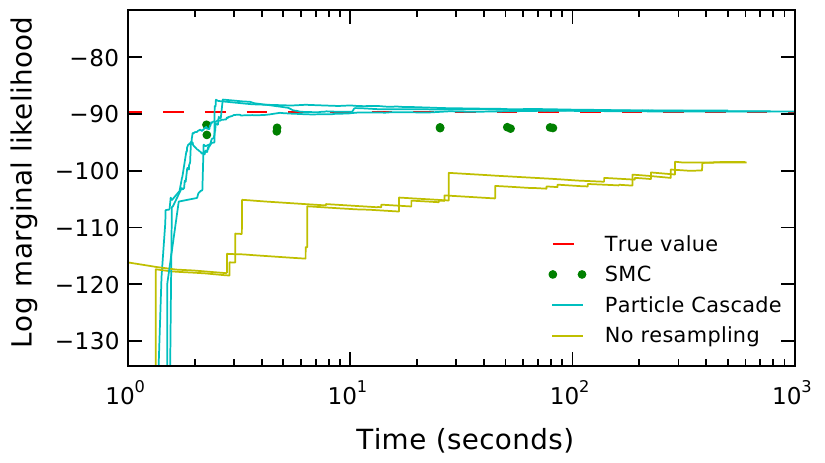}
\caption{(top) Comparative convergence rates between SMC alternatives including our new algorithm,
and (bottom) estimation of marginal likelihood, by time. 
Results are shown for
(left) the hidden Markov model, and (right) the linear Gaussian state space model.}
\label{fig:wallclock-time}
\end{center}
\vspace{-.5em}
\end{figure}

Although values will be implementation-dependent, we are ultimately interested not in per-sample efficiency but rather in our rate of convergence over time.
We record wall clock time for each algorithm for both of these models; the results for convergence of our estimates of values and marginal likelihood are shown in Fig.~\ref{fig:wallclock-time}.
These particular experiments were all run on Amazon EC2, in an 8-core environment with Intel Xeon E5-2680 v2 processors.
The particle cascade provides a much faster and more accurate estimate of the marginal likelihood than the competing methods, in both models.
Convergence in estimates of values is quick as well, faster than the iCSMC approach.
We note that for very small numbers of particles, running a simple particle filter is faster than the particle cascade, despite the blocking nature of the resampling step.
This is due to the overhead incurred by the particle cascade in sending an initial flurry of $\rho_0$ particles into the system before we see any particles progress to the end; this initial speed advantage diminishes as the number of samples increases.
Furthermore, in stark contrast to the simple SMC method, there are no barriers to drawing more samples from the particle cascade indefinitely.
On this fixed hardware environment, our implementation of SMC, which aggressively parallelizes all forward particle simulations, exhibits a dramatic loss of performance as the number of particles increases from $10^4$ to $10^5$, to the point where simultaneously running $10^5$ particles is simply not possible in a feasible amount of time.

%

\begin{figure}[t]
\begin{center}
\includegraphics[width=.7\textwidth]{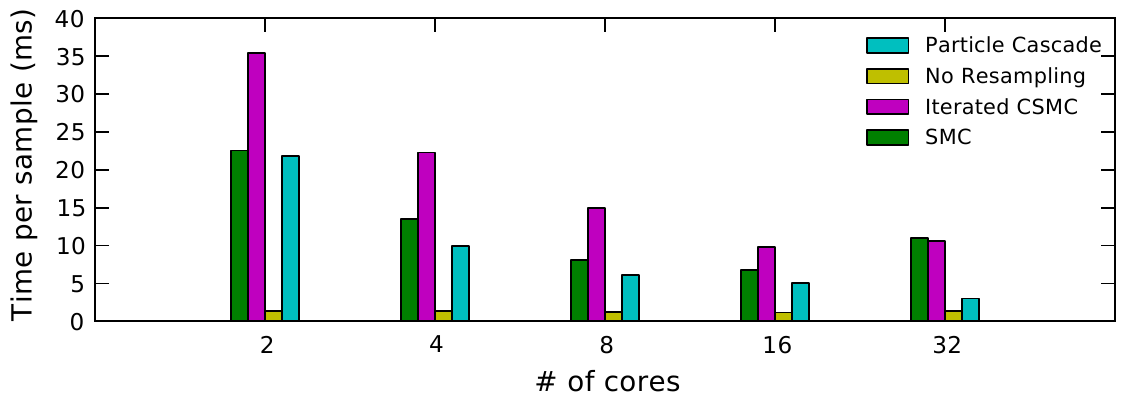}
\vspace{-1em}
\caption{Average time to draw a single complete particle on a variety of machine architectures. 
Queueing rather than blocking at each observation improves performance, and appears to improve relative performance even more as the available compute resources increase.
Note that we see only average time per sample, here; this is not a measure of statistical efficiency.
The high speed of the non-resampling algorithm is not sufficient to make it competitive with the other approaches.
}
\label{fig:time-to-compute-n-particles}
\end{center}
\end{figure}

We are also interested in how the particle cascade scales up to larger hardware, or down to smaller hardware.
A comparison across 5 different hardware configurations is shown in Fig.~\ref{fig:time-to-compute-n-particles}.



\section{Discussion}
\label{sec:discussion}

The particle cascade has broad applicability, appropriate for all SMC and particle filtering inference applications.
For example, constructing an appropriate sequence of densities for SMC is possible in arbitrary probabilistic graphical models, including undirected graphical models; see e.g.~the sequential decomposition approach of \cite{naesseth2014sequential}.
%
We are particularly motivated by the SMC-based probabilistic programming systems that have recently appeared in the literature \cite{wood2014anglican,paige14icml}. 
In both references it was suggested that primary performance bottleneck in their inference algorithms was barrier synchronization, something we have done away with entirely.  
What is more, while particle MCMC methods are particularly appropriate when there is a clear boundary that can be exploited between between parameters of interest and nuisance state variables, in a growing number of applications in the probabilistic programming and SMC communities, parameters values are generated as part of the state trajectory itself, leaving no explicitly denominated latent parameter variables per se.  
The particle cascade is particularly relevant to such approaches.

Finally, an attractive property of this algorithm is that it yields an unbiased estimate of the marginal likelihood, and thus can be plugged directly into PIMH, SMC$^2$ \cite{chopin2013smc2}, and other so-called pseudomarginal methods.

%
%
%
%
%


\subsubsection*{Acknowledgments}

Yee Whye Teh's research leading to these results has received funding
from EPSRC (grant EP/K009362/1) and the ERC under the EU's FP7
Programme (grant agreement no. 617411).
Frank Wood is supported under DARPA PPAML.  
This material is based on research sponsored by DARPA through the U.S. Air Force Research Laboratory under Cooperative Agreement number FA8750-14-2-0004.  The U.S. Government is authorized to reproduce and distribute reprints for Governmental purposes notwithstanding any copyright notation heron.
The views and conclusions contained herein are those of the authors and should be not interpreted as necessarily representing the official policies or endorsements, either expressed or implied, of DARPA, the U.S. Air Force Research Laboratory of the U.S. Government.

%

\subsubsection*{References}
\begingroup
\renewcommand{\section}[2]{}%
\bibliographystyle{unsrtnat}
\bibliography{refs}

\begin{thebibliography}{11}
\providecommand{\natexlab}[1]{#1}
\providecommand{\url}[1]{\texttt{#1}}
\expandafter\ifx\csname urlstyle\endcsname\relax
  \providecommand{\doi}[1]{doi: #1}\else
  \providecommand{\doi}{doi: \begingroup \urlstyle{rm}\Url}\fi

\bibitem[Whiteley et~al.(2013)Whiteley, Lee, and Heine]{whiteley2013}
Nick Whiteley, Anthony Lee, and Kari Heine.
\newblock {On the role of interaction in sequential Monte Carlo algorithms}.
\newblock \emph{arXiv preprint arXiv:1309.2918}, 2013.

\bibitem[Andrieu et~al.(2010)Andrieu, Doucet, and
  Holenstein]{andrieu2010particle}
Christophe Andrieu, Arnaud Doucet, and Roman Holenstein.
\newblock {Particle Markov chain Monte Carlo methods}.
\newblock \emph{Journal of the Royal Statistical Society: Series B (Statistical
  Methodology)}, 72\penalty0 (3):\penalty0 269--342, 2010.

\bibitem[Crisan et~al.(1999)Crisan, Moral, and Lyons]{crisan1999filtering}
D.~Crisan, P.~Del Moral, and T.~Lyons.
\newblock Discrete filtering using branching and interacting particle systems.
\newblock \emph{Markov Process. Related Fields}, 5\penalty0 (3):\penalty0
  293--318, 1999.

\bibitem[Murray et~al.(2014)Murray, Lee, and Jacob]{murray2014parallel}
Lawrence~M. Murray, Anthony Lee, and Pierre~E. Jacob.
\newblock Parallel resampling in the particle filter.
\newblock \emph{arXiv preprint arXiv:1301.4019}, 2014.

\bibitem[Jun and Bouchard-C\^ot\'e(2014)]{jun14icml}
Seong-Hwan Jun and Alexandre Bouchard-C\^ot\'e.
\newblock Memory (and time) efficient sequential monte carlo.
\newblock In \emph{Proceedings of the 31st international conference on Machine
  learning}, 2014.

\bibitem[Douc et~al.(2005)Douc, Capp\'e, and Moulines]{douc05comparisonof}
Randal Douc, Olivier Capp\'e, and Eric Moulines.
\newblock Comparison of resampling schemes for particle filtering.
\newblock In \emph{In 4th International Symposium on Image and Signal
  Processing and Analysis (ISPA)}, pages 64--69, 2005.

\bibitem[Moral(2004)]{delmoral04feynman}
Pierre~Del Moral.
\newblock \emph{Feynman-Kac Formulae -- Genealogical and Interacting Particle
  Systems with Applications}.
\newblock Probability and its Applications. Springer, 2004.

\bibitem[Nasseth et~al.(2014)Nasseth, Lindsten, and
  Sch\:on]{naesseth2014sequential}
Christian~A. Nasseth, Fredrik Lindsten, and Thomas~B. Sch\:on.
\newblock Sequential monte carlo methods for graphical models.
\newblock \emph{arXiv preprint arXiv:1402.0330}, 2014.

\bibitem[Wood et~al.(2014)Wood, van~de Meent, and Mansinghka]{wood2014anglican}
Frank Wood, Jan~Willem van~de Meent, and Vikash Mansinghka.
\newblock A new approach to probabilistic programming inference.
\newblock In \emph{Proceedings of the 17th International conference on
  Artificial Intelligence and Statistics}, 2014.

\bibitem[Paige and Wood(2014)]{paige14icml}
Brooks Paige and Frank Wood.
\newblock A compilation target for probabilistic programming languages.
\newblock In \emph{Proceedings of the 31st international conference on Machine
  learning}, 2014.

\bibitem[Chopin et~al.(2013)Chopin, Jacob, and
  Papaspiliopoulos]{chopin2013smc2}
Nicolas Chopin, Pierre~E Jacob, and Omiros Papaspiliopoulos.
\newblock Smc2: an efficient algorithm for sequential analysis of state space
  models.
\newblock \emph{Journal of the Royal Statistical Society: Series B (Statistical
  Methodology)}, 75\penalty0 (3):\penalty0 397--426, 2013.

\end{thebibliography}
\endgroup

\end{document}